\documentclass[preprintnumbers,article,amsmath,amssymb,floatfix,10pt,prd,onecolumn, superscriptaddress,nofootinbib]{revtex4}
\usepackage[colorlinks=true, pdfstartview=FitV, linkcolor=blue, citecolor=red, urlcolor=magenta]{hyperref}
\usepackage{bbm}
\usepackage{amsfonts}
\usepackage{mathrsfs}
\usepackage{latexsym}
\usepackage{epsfig}
\usepackage{epstopdf}
\usepackage{epstopdf}
\usepackage{graphicx}
\usepackage{amssymb}
\usepackage{amsmath}
\usepackage{dcolumn}
\usepackage{bm}
\usepackage{color}
\usepackage{comment}
\usepackage{xcolor}
\begin{document}
\title{\bf New embedded wormhole solutions in Ricci inverse gravity}

\author{G. Mustafa}
\email{gmustafa3828@gmail.com}\affiliation{Department of Physics, Zhejiang Normal University, Jinhua 321004, People's Republic of China}

\author{Faisal Javed}
\email{faisaljaved.math@gmail.com} 
\affiliation{Department of Physics, Zhejiang Normal University, Jinhua 321004, People's Republic of China}

\author{S. K. Maurya}
\email{sunil@unizwa.edu.om}\affiliation{Department of Mathematical and Physical Sciences, College of Arts and Sciences, University of Nizwa, Nizwa, Sultanate of Oman}

\author{Abdelghani Errehymy}
\email{abdelghani.errehymy@gmail.com}\affiliation{Astrophysics Research Centre, School of Mathematics, Statistics and Computer Science, University of KwaZulu-Natal, Private Bag X54001, Durban 4000, South Africa}

\begin{abstract}
In this letter, we obtain two new embedded WH solutions by using the class-I approach in the background of newly fourth-order Ricci inverse gravity. We show that the combination of these newly calculated shape functions and Ricci inverse gravity provides us with the possibility of obtaining traversable wormholes. All the required wormhole properties are discussed, along with flaring out and flatness conditions. The embedded diagrams within the scope of upper and lower universes are provided under the effect of both newly calculated embedded shape functions. All the energy conditions are explored with valid and negative regions. The presence of exotic matter is confirmed due to the negative region in all the energy conditions, specifically in the null energy condition. The Doppler effect through the red-blue shifts function is also discussed. Several key findings from the current research are described that demonstrate the validity of these wormhole solutions in Ricci inverse gravity. 
\end{abstract}

\maketitle

\date{\today}


\section{Introduction}

A major challenge in contemporary cosmology is the ongoing phenomenon of the universe's accelerating expansion. There is compelling empirical evidence supporting this occurrence \cite{hf23,hf24,hf25}. One potential explanation for comprehending this rapidly progressing stage is the inclusion of a novel and unconventional substance known as dark energy.
On enormous cosmic scales, the negative pressure of this component makes gravity behave repulsively \cite{hf26}. The most well-known potential dark energy possibility is the cosmological constant $\Lambda$. The $\Lambda$CDM model has been developed, taking this component and assuming the presence of dark matter. Despite being a widely accepted model that effectively explains the observed data, still it has issues such as well-known cosmological constant problem \cite{hf27}. Modifications to Einstein's general relativity are suggested as an additional explanation for the cosmic data \cite{hf28,hf29}. Alternate theories of general relativity can be constructed most straightforwardly by either incorporating an extra term into the Einstein-Hilbert Lagrangian or by adjusting the structure of the Lagrangian, which would include revising the Ricci scalar itself. A relatively new family of intriguing fourth-order modified gravity models is provided by the Ricci-inverse gravity framework \cite{Amendola}. Ricci-inverse gravity is a relatively new model that has sparked a lot of attention among the multiple modified gravity theories that are now available in the scientific literature. The function of the anti-curvature scalar $A$ and the Ricci scalar $R$ generalizes the Einstein-Hilbert action. One must pay attention to the fact that the Ricci scalar $R$ inverse is not the anti-curvature scalar $A$. A study on modified gravity actions with negative or positive anti-curvature scalar powers was conducted by Amendola et al. \cite{Amendola}. The researchers have shown that cosmic trajectories cannot transition smoothly from a decelerated phase to a substantially accelerated phase. They have also developed a universal theory that prohibits such transitions in these mathematical models.
In the framework of inverse Ricci gravity models, Das et al. \cite{RI2} showed that the reduced action technique cannot overcome the no-go theorem. In addition, they deliberated on the implications of their discoveries for the cosmology of the early cosmos.  Souzaa and Santos \cite{RI3} computed two distinct axially symmetric spacetimes exhibiting causality violation, demonstrating that Ricci-inverse gravity allows for the presence of closed time-like curves. Tuan Q. Do \cite{RI4} introduced singularity-free solutions for both anisotropically as well as isotropically inflating Universes.  The theory has been investigated in the framework of cosmic structures, specifically concerning the stability of the Sub-Horizon non-relativistic Weak-Field limit for matter distributions that are static and spherically symmetric and immersed in a de Sitter cosmology \cite{hf4}. These gravitational theories have been utilized in several research that have been carried out in recent years. Examples of applications include the investigation of the no-go theorem for inflation \cite{hf1}, the study of cosmic structure \cite{hf2}, and the exploration of anisotropic star structures \cite{hf3}, amongst other applications. Jawad and his coauthor \cite{RI5}  conducted a recent study on the matter-antimatter asymmetry phenomena. They specifically focused on baryogenesis within the framework of Ricci inverse gravity. In our current analysis, we explore novel aspects of Wormhole (WH) geometry in the setting of Ricci inverse gravity, drawing inspiration from studies on WH solutions and Ricci inverse gravity.

An intriguing idea in General Relativity is the Einstein-Rosen bridge, which is building a structure like a bridge to connect places inside spacetime or distinct spacetimes. The notion of a WH was first introduced by Flamm \cite{52t} in 1916, and Misner and Wheeler popularised the concept of a WH in 1957. The possibility of instability in an Einstein-Rosen bridge-type WH was highlighted by Fuller and Wheeler \cite{53t} in 1962. The idea was that something slower than light could not possibly travel between two regions of the same cosmos since the connection between them could collapse quickly. The presence of WHs raises specific questions, yet research on these structures has significantly advanced our understanding of modern astrophysics. Morris and Thorne made a landmark discovery in 1988 \cite{54t} when they introduced traversable WHs that are static as well as spherically symmetric and supported through exotic matter located at the WH throat. The WH's throat is filled with exotic materials to counteract the gravitational forces that could cause the WH to collapse quickly. Because exotic matter can produce negative pressure and go against the Null Energy Condition (NEC), it is taken into consideration for its ability to function as a counteracting force against the gravitational attraction \cite{55t}. A crucial indicator for the possibility of a traversable WH is a violation of energy constraints, most notably the NEC, which may be associated with exotic matter, at least in the vicinity of the throat region \cite{56t,57t,58t}. It is important to remember that, despite being thoroughly researched in various theories, WHs are still purely theoretical phenomena that still need to be discovered using astronomical methods in an experiment.

A significant number of research on WH solutions and their connection to current astrophysical occurrences focuses on modified theories of gravity. Reducing the requirement for stabilizing WH geometry, such hypotheses allow the investigation of various WH geometries with conventional matter rather than exotic matter \cite{60t,68t}. Researchers have examined traversable WH geometries within the context of modified gravity theories, such as $f(R)$ gravity with specific forms of functions \cite{68f,68y} and equations of the state \cite{69t}. In addition, studies have been conducted on traversable WH solutions in $f(G)$ gravity \cite{70t}, $f(R, \mathcal{T})$ gravity \cite{71t}. Furthermore, alongside Riemannian geometry, several geometric frameworks have also contributed to discussions regarding the feasibility of traversable WH solutions and their inherent physical characteristics. WH solutions are investigated in the context of symmetric teleparallel gravity, specifically in the frameworks of $f(Q)$ and $f(Q, \mathcal{T})$ gravities \cite{75t,76t}, teleparallel geometry with $f(T)$ gravity  \cite{73t}, and $f(T, \mathcal{T})$ gravity \cite{74t,75g}. Garattini et al. \cite{77t}  studied traversable WHs in the field of non-commutative geometry. In recent studies on Fislerian and Finsler-Randers geometry \cite{78t,79t}, 
In a power-law $f(R)$ model, traversable WH solutions were studied by Capozziello et al. \cite{Capozziello:2020zbx} utilizing two special shape functions. A convincing substitute for dealing with the problem of exotic matter is provided by modified theories. A wide range of traversable WH geometries within the context of alternative gravitational theories have been thoroughly investigated in the currently available literature \cite{ovgun2018,cap1,cap2,cap3,cap4,cap5}.

As previously stated, the focus of this letter will be on WH solutions in the framework of inverse Ricci gravity. The field equations of fourth-order gravity will be deeply discussed in Sec. 2. The geometry of the embedded WH spacetime along embedded diagrams should be presented in Sec. 3. Sec. 4 calculates and presents the energy conditions, including valid and negative regions. Sec. 5 deals with the Doppler effect through the Red-blue shift function. Finally, in Sec. 6, we analyze the physics behind our findings and provide our conclusive remarks.

\section{Ricci Inverse Gravity}

In this section, we explore Ricci inverse gravity and calculate the final version of field equations. The Ricci inverse gravity starts from inverse tensor of $R_{\varsigma\tau}$, which is expressed as
\begin{equation} \label{14}
A^{\varsigma\tau}R_{c}=\delta^{\varsigma}_{c}.
\end{equation}
In the above equation, $A^{\varsigma\tau}$ mentions the anti-curvature tensor. The extended action for inverse Ricci gravity was described by \cite{Amendola}, and it is read as:
\begin{equation} \label{15}
S=\int \sqrt{-g}d^4x(\gamma A+R),
\end{equation}
where $A$ is a trace of $A^{\varsigma\tau}$, which is defined as
\begin{equation} \label{16}
A^{\varsigma\tau}=R^{-1}_{\varsigma\tau}.
\end{equation}
From Eq. (\ref{16}), one can calculate the following expression
\begin{equation} \label{17}
\delta A^{\varsigma \varrho}=-A^{\varsigma j}\left(\delta R_{\tau \sigma}\right) A^{\sigma \varrho} .
\end{equation}
Further, Eq. (\ref{15}) helps us to calculate the following relation
\begin{eqnarray}\label{18}
\delta S && =\int d^4 x\left(A \delta \sqrt{-g}+\sqrt{-g} A^{\varsigma v} \delta g_{\varsigma \tau}+\sqrt{-g} g_{\varsigma \tau} \delta A^{\varsigma \tau}\right), \\\label{19}
&& =\int d^4 x \sqrt{-g}\left(\frac{1}{2} A g^{\varsigma v} \delta g_{\varsigma \tau}+A^{\varsigma v} \delta g_{\varsigma \tau}+g_{\varsigma \tau} \delta A^{\varsigma \tau}\right),
\end{eqnarray}
and 
\begin{equation} \label{20}
\delta R_{\epsilon \beta}=\nabla_\rho \delta \Gamma_{\beta \epsilon}^\rho-\nabla_\beta \delta \Gamma_{\rho \epsilon}^\rho,
\end{equation}
one can get
\begin{eqnarray}\label{21}
\delta A^{\varsigma j} && =-A^{\varsigma \epsilon}\left(\nabla_\rho \delta \Gamma_{\beta \epsilon}^\rho-\nabla_\beta \delta \Gamma_{\rho \epsilon}^\rho\right) A^{\beta \tau}, \\\label{22}
&& =-\frac{1}{2} A^{\varsigma \epsilon}\left(g^{\rho \lambda} \nabla_\rho\left(\nabla_\epsilon \delta g_{\beta \lambda}+\nabla_\beta \delta g_{\lambda \epsilon}-\nabla_\lambda \delta g_{\epsilon \beta}\right)-g^{\rho \lambda} \nabla_\beta\left(\nabla_\epsilon \delta g_{\rho \lambda}+\nabla_\rho \delta g_{\lambda \epsilon}-\nabla_\lambda \delta g_{\epsilon \rho}\right)\right) A^{\beta \tau}, \\\label{23}
&& =-\frac{1}{2} A^{\varsigma \epsilon} g^{\rho \lambda}\left(\nabla_\rho \nabla_\epsilon \delta g_{\beta \lambda}-\nabla_\rho \nabla_\lambda \delta g_{\epsilon \beta}-\nabla_\beta \nabla_\epsilon \delta g_{\rho \lambda}+\nabla_\beta \nabla_\lambda \delta g_{\epsilon \rho}+\left[\nabla_\beta, \nabla_\rho\right] \delta g_{\lambda \epsilon}\right) A^{\beta \tau}.
\end{eqnarray}
Now, by employing the integration-by-parts approach, we have
\begin{eqnarray}\label{24}
g_{\varsigma \tau} \delta A^{\varsigma \tau} && =-\frac{1}{2} g_{\varsigma \tau} g^{\rho \lambda}\left(\delta g_{\beta \lambda} \nabla_\epsilon \nabla_\rho\left(A^{\varsigma \epsilon} A^{\beta v}\right)-\delta g_{\epsilon \beta} \nabla_\lambda \nabla_\rho\left(A^{\varsigma \epsilon} A^{\beta v}\right)-\delta g_{\rho \lambda} \nabla_\epsilon \nabla_\beta\left(A^{\varsigma \epsilon} A^{\beta v}\right)\right.\nonumber\\&&\left.+\delta g_{\epsilon \rho} \nabla_\lambda \nabla_\beta\left(A^{\varsigma \epsilon} A^{\beta v}\right)\right), \\\label{25}
&& =\frac{1}{2} \delta g_{\iota \kappa}\left(-2 g^{\rho \iota} \nabla_\epsilon \nabla_\rho A^{\varsigma \epsilon} A_\varsigma^\kappa+\nabla^2\left(A^{\varsigma \iota} A_\varsigma^\kappa\right)+g^{\iota \kappa} \nabla_\epsilon \nabla_\beta\left(A^{\varsigma \epsilon} A_\varsigma^\beta\right)\right).
\end{eqnarray}
So finally, we have the following relation
\begin{equation} \label{26}
\delta g_{\varsigma \tau}\left(\frac{1}{2} A g^{\varsigma \tau}+A^{\varsigma \tau}+\frac{1}{2}\left(-2 g^{\rho \varsigma} \nabla_\epsilon \nabla_\rho A^{\sigma \epsilon} A_\sigma^\tau+\nabla^2\left(A^{\sigma \varsigma} A_\sigma^\tau\right)+g^{\varsigma \tau} \nabla_\epsilon \nabla_\beta\left(A^{\sigma \epsilon} A_\sigma^\beta\right)\right)\right).
\end{equation}
On applying a variation of the Lagrangian, one can get the following relation
\begin{equation} \label{27}
\delta g^{\varsigma \tau}\left(-\frac{1}{2} R g_{\varsigma \tau}+R_{\varsigma \tau}\right)=-\delta g_{\varsigma \tau}\left(-\frac{1}{2} R g^{\varsigma \tau}+R^{\varsigma \tau}\right).
\end{equation}
By taking the procedure straightway from \cite{Amendola}, hence the respective improved version of field equations yield
\begin{equation}\label{28}
R^{\varsigma \tau}-\frac{1}{2}Rg^{\varsigma \tau}-\gamma A^{\varsigma \tau}-\frac{1}{2}\gamma A g^{\varsigma \tau}+\frac{\gamma}{2}\left(2g^{\varrho \varsigma}\nabla_{\epsilon}\nabla_{\varrho}A^{\epsilon}_{\sigma}A^{b\sigma}-\nabla^2A^{\varsigma}_{\sigma}
A^{\tau\sigma}-g^{\varsigma \tau}\nabla_{\epsilon}\nabla_{\varrho}A^{\epsilon}_{\sigma}A^{\varrho\sigma}\right)={\mathcal{T}}^{\varsigma \tau},
\end{equation}
where $A^{\epsilon}_{\sigma}A^{b\sigma}=A^{\epsilon\varrho}g_{\varrho\sigma}A^{\sigma b}=A^{\epsilon\varrho}A^{b}_{\varrho}=A^{\epsilon\sigma}A^{b}_{\sigma}
=A^{b}_{\sigma}A^{\epsilon\sigma}$ with $8\pi G=1$.

\subsection{WH geometry and Fiedl equations of Ricci Inverse Gravity}
To explore inverse Ricci gravity in the background of WH geometry. Firstly, we consider the stress tensor with anisotropic matter content, expressed as
\begin{equation} \label{29}
{\mathcal{T}}_{ij}=(\rho+p_{t})u_{i}u_{j}-p_{t}g_{ij}+(p_{r}-p_{t})v_{i}v_{j}.
\end{equation}
Here, $u_{i}=e^{\frac{b}{2}}\delta_{i}^{0}$ and $v_{i}=e^{\frac{\lambda}{2}}\delta_{i}^{1}$. For the line element choice, we consider the spherically symmetric metric, which is defined as
\begin{equation} \label{30}
ds^{2}=e^{a(r)}dt^{2}-e^{b(r)} dr^{2}-r^{2} d\theta^{2}-r^{2}\sin^{2}\theta d\phi^{2}.
\end{equation}
The Morris and Thorne WH geometry via spherically symmetric within the scope of Schwarzschild coordinates \cite{55t} is expressed as
\begin{equation}\label{31}
e^{a(r)}=e^{2\Phi(r)}, \;\;\;\;\;\;\;\;\;\;\;\;e^{b(r)}=\left(1-\frac{\mathcal{S}(r)}{r}\right)^{-1},
\end{equation}
where $\Phi(r)$ and $\mathcal{S}(r)$ are defining red-shift and shape function respectively. 
Now, by plugging the Eqs. (\ref{29}), (\ref{30}), and (\ref{31}) in Eq. (\ref{28}), we get the following modified field equations:
\begin{eqnarray}
\rho &=&\frac{(r-\mathcal{S}(r)) \left(-\frac{2 \gamma  \left(\rho _1-\rho _2\right) r^3}{(r-\mathcal{S}(r))^2}+\frac{4 \gamma  \rho _{13} r^2}{\rho _{12}}-\frac{\mathcal{S}'(r)}{r^2-r \mathcal{S}(r)}\right)}{r}, \label{32}\\
p_r&=&-4 \gamma  r \left(\frac{r \left(p_6 r^4-p_9 r^3 \mathcal{S}(r)+p_5 r^2 \mathcal{S}(r)^2-p_4 r \mathcal{S}(r)^3+p_3 \mathcal{S}(r)^4\right)}{p_{10}^4}+\frac{p_2}{p_1^3}\right),\label{33}\\
p_t&=&-\frac{r \left(-8 \gamma  \left(p_{12}+p_{13}\right) r^2+8 \gamma  p_{17} r^2+p_{11}\right)}{4 (r-\mathcal{S}(r))}\label{34},
\end{eqnarray}
where $p _i$, $\{i=1,...,17\}$ and $\rho _i$, $\{i=1,...,13\}$ are provided in the Appendix. 

\section{Embedding Class I condition and new WH solutions}

In this research, we will use embedded class-1 spacetime within the scope of Karmarkar condition \cite{em1} to obtain a new class of WH solutions. One of the most important aspects of the current analysis is the Karmarkar condition. The embedded class-1 solution of Riemannian space is the requirement for the fundamental construction of the Karmarkar condition. Eisenhart  \cite{em2} suggested a necessary and suitable condition for the embedded class-1 solution, which is dependent upon the Riemann curvature tensor, $R_{a b \alpha \beta}$, and a symmetric tensor of the second order, $B_{a b}$. To be classified as a class-1 spacetime, any spherically symmetric spacetime, whether static or non-static, must satisfy the following necessary and sufficient conditions:

\begin{itemize}
  \item Establishing the relation below is necessary for a system with a symmetric tensor of the second order, $B_{a b}$:
\end{itemize}
\begin{equation}
    \text{Gauss equation}:\;\;\;\; R_{a b \alpha \beta}=\epsilon\left(B_{a \alpha} B_{v \beta}-B_{a \beta} B_{b \alpha}\right),
\end{equation}
where $\epsilon$ provide the space-like or time-like manifold with $\epsilon=+1$ and $\epsilon=-1$ respectively.
\begin{itemize}
  \item The above mentioned symmetric tensor $B_{a b}$ should be meet the following relation:
\end{itemize}
\begin{eqnarray}
    \text { Codazzi's equation}:\;\;\;\;\nabla_{\alpha} B_{a b}-\nabla_{b} B_{a \alpha}=0. \quad 
\end{eqnarray}
Now, the Riemannian components within Schwarzschild's coordinates $(t, r, \theta, \phi \equiv$ $0,1,2,3$ ) are calculated as:
\begin{eqnarray*}
& R_{r t r t}=-e^{a}\left(\frac{a^{\prime \prime}}{2}-\frac{b^{\prime} a^{\prime}}{4}+\frac{a^{\prime 2}}{4}\right) ; \quad R_{r \theta r \theta}=-\frac{r}{2} b^{\prime} ; 
 R_{\theta \phi \theta \phi}=-\frac{r^{2} \sin ^{2} \theta}{e^{b}}\left(e^{b}-1\right) ; \quad R_{r \phi \phi t}=0,\\ 
& R_{\phi t \phi t}=-\frac{r}{2} a^{\prime} e^{a-b} \sin ^{2} \theta ; \quad R_{r \theta \theta t}=0 .
\end{eqnarray*}
Now, the Gauss equation reads as
\begin{eqnarray}
    b_{t r} b_{\phi \phi} = R_{r \phi t \phi}=0 ; \quad b_{t r} b_{\theta \theta}=R_{r \theta t \theta}=0, \\
    b_{t t} b_{\phi \phi}=R_{t \phi t \phi} ; \quad b_{t t} b_{\theta \theta} = R_{t \theta t \theta} ; \quad b_{r r} b_{\phi \phi}=R_{r \phi r \phi},\\
    b_{\theta \theta} b_{\phi \phi} = R_{\theta \phi \theta \phi} ; \quad b_{r r} b_{\theta \theta}=R_{r \theta r \theta} ; \quad b_{t t} b_{r r}=R_{t r t r}.
\end{eqnarray}
From the previous set of relations, we have the following expressions
\begin{eqnarray}\label{Rie1}
    \left(b_{t t}\right)^{2}=\frac{\left(R_{t \theta t \theta}\right)^{2}}{R_{\theta \phi \theta \phi}} \sin ^{2} \theta, \quad\left(b_{r r}\right)^{2}=\frac{\left(R_{r \theta r \theta}\right)^{2}}{R_{\theta \phi \theta \phi}} \sin ^{2} \theta,\\ \label{Rie2}
    \left(b_{\theta \theta}\right)^{2}=\frac{R_{\theta \phi \theta \phi}}{\sin ^{2} \theta}, \quad\left(b_{\phi \phi}\right)^{2}=\sin ^{2} \theta R_{\theta \phi \theta \phi}.
\end{eqnarray}
One can get the important connection in Riemann components in the framework of Eq.(\ref{Rie1}) and Eq. (\ref{Rie2}) as
\begin{eqnarray}
    R_{t \theta t \theta} R_{r \phi r \phi}=R_{t r t r} R_{\theta \phi \theta \phi}.
\end{eqnarray}
It is noteworthy to mention that Codazzi's equation is provided by the equations above. An important point to note is that the symmetric tensor $B_{ab}$ has the following equation in the case of a non-static spherically symmetric spacetime:
\begin{eqnarray}
    b_{t r} b_{\theta \theta}=R_{r \theta t \theta} \quad \text{and} \quad b_{t t} b_{r r}-\left(b_{t r}\right)^{2}=R_{t r t r}.
\end{eqnarray}
In the above equation, we have the following relation: $\left(b_{t r}\right)^{2}=\sin ^{2} \theta\left(R_{r \theta t \theta}\right)^{2} / R_{\theta \phi \theta \phi}$. In this scenario, the embedding Class I condition has the following form:
\begin{eqnarray}
    R_{t \theta t \theta} R_{r \phi r \phi}=R_{t r t r} R_{\theta \phi \theta \phi}+R_{r \theta t \theta} R_{r \phi t \phi}.
\end{eqnarray}
The above equation is also known as the Karmarkar condition. However, in our specific case, the condition will be comparable to that of the static, spherically symmetric metric. The preceding requirement is a necessary and sufficient condition for classifying a spacetime as Class-I. By plugging the Riemann components into the condition mentioned above, one can get the differential equation:
\begin{eqnarray}
    2 \frac{a^{\prime \prime}}{a^{\prime}}+a^{\prime}=\frac{b^{\prime} e^{b}}{e^{b}-1}.
\end{eqnarray}
On integrating the above second-order differential equation, one can produce the following relationship between gravitational potentials:
\begin{eqnarray}\label{final}
    e^{b}=1+A a^{\prime 2} e^{a}, 
\end{eqnarray}
where $A$ is the integration constant. Now, due to the embedded class one solution, we must consider a specific form of redshift function as 
\begin{equation}\label{red1}
 a(r)=2\Phi(r)=\frac{\zeta _2}{r^{\zeta _1}},   
\end{equation}
where $\zeta _1$ and $\zeta _2$ are constants. The chosen redshift function satisfies the flatness condition. By adopting the procedure, which is already reported in \cite{em3} along Eq. (\ref{31}) and Eq. (\ref{final}), one can obtain the following relation for the shape function as:
\begin{eqnarray}\label{shape1}
\mathcal{S}(r)=\frac{r (C_{1}-r_{0}) e^{\zeta _2 r^{-\zeta _1}} r_{0}^{2 \zeta _1+2}}{-C_{1} r^{2 \zeta _1+2} e^{\zeta _2 r_{0}^{-\zeta _1}}+C_{1} e^{\zeta _2 r^{-\zeta _1}} r_{0}^{2 \zeta _1+2}-e^{\zeta _2 r^{-\zeta _1}} r_{0}^{2 \zeta _1+3}}+C_{1},\;\;\; 0<C_{1}\leq r_{0}.
\end{eqnarray}
In the current analysis, the above shape function by Eq. (\ref{shape1}) should be known as Model-I. For a more general case, we consider another type of redshift function, which also satisfies the flatness condition, and it is defined as: 
\begin{equation}\label{red2}
 a(r)=2\Phi(r)=-2 r^{\chi _1} \omega ^{\chi _1}-\frac{2 \chi _2}{r}.   
\end{equation}
In the above Eq. (\ref{red2}), $\chi _1$ and $\chi _2$ are constants. Similarly, by adopting the procedure from \cite{em3} along Eq. (\ref{31}) and Eq. (\ref{final}), one can get another general form of embedded shape function, which is calculated as:
\begin{eqnarray}\label{shape2}
\mathcal{S}(r)=\frac{r r_{0}^4 S_1^2 (C_{2}-r_{0}) e^{2 r_{0}^{\chi _1} \omega ^{\chi _1}+\frac{2 \chi _2}{r_{0}}}}{C_{2} \left(r_{0}^4 S_1^2 e^{2 r_{0}^{\chi _1} \omega ^{\chi _1}+\frac{2 \chi _2}{r_{0}}}-r^4 S_1^2 e^{2 r^{\chi _1} \omega ^{\chi _1}+\frac{2 \chi _2}{r}}\right)-r_{0}^5 S_1^2 e^{2 r_{0}^{\chi _1} \omega ^{\chi _1}+\frac{2 \chi _2}{r_{0}}}}+C_{2},\;\;\; 0<C_{2}\leq r_{0},
\end{eqnarray}
where 
\begin{equation*}
S_1=\chi _2-\chi _1 r^{\chi _1+1} \omega ^{\chi _1}.
\end{equation*}
\begin{figure}
\centering \epsfig{file=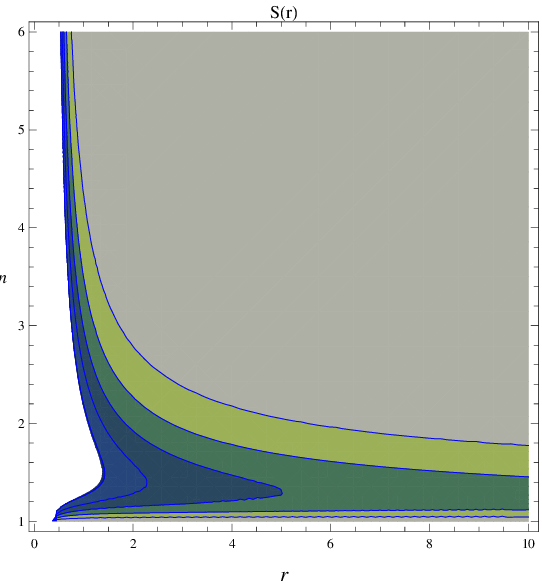, width=.26\linewidth,
height=1.7in}\epsfig{file=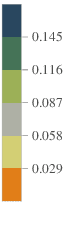, width=.05\linewidth,
height=1.7in} \epsfig{file=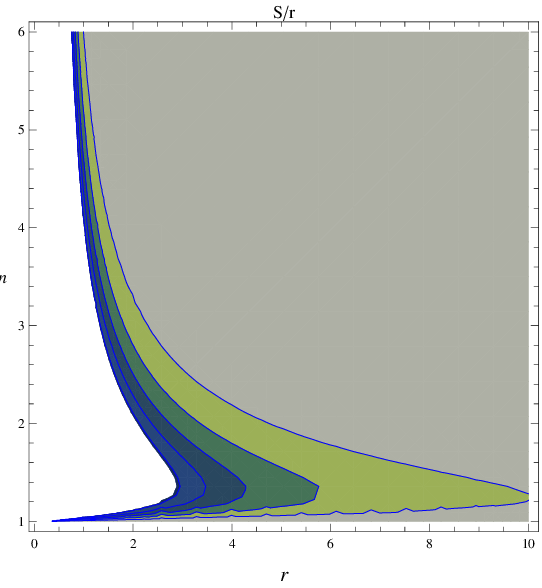, width=.26\linewidth,
height=1.7in}\epsfig{file=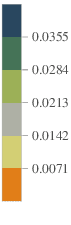, width=.05\linewidth,
height=1.7in} \epsfig{file=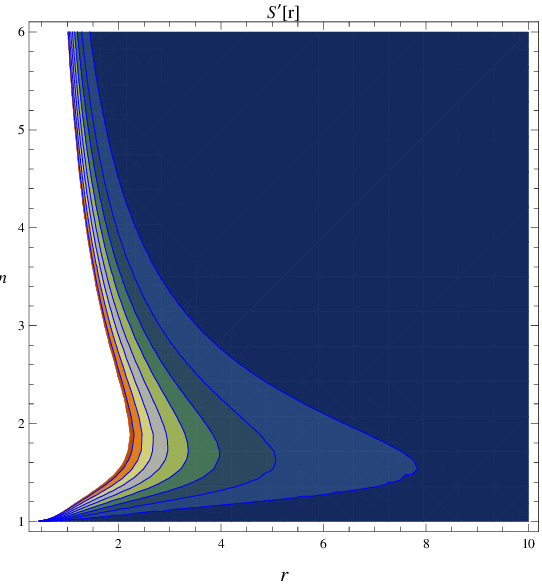, width=.26\linewidth,
height=1.7in}\epsfig{file=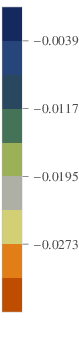, width=.05\linewidth,
height=1.7in}
\caption{\label{F1} shows the shape function properties for model-I with $S(r)$(left), $\frac{S(r)}{r}$ (middle) and $\frac{dS}{dr}$ (right).}
\end{figure}

\begin{figure}
\centering \epsfig{file=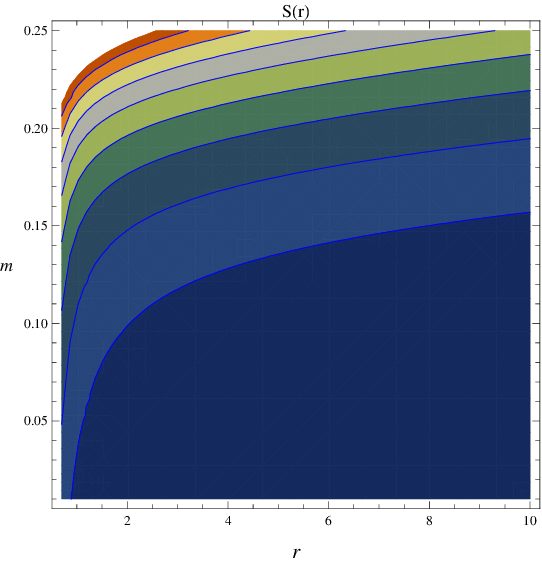, width=.26\linewidth,
height=1.7in}\epsfig{file=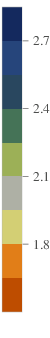, width=.05\linewidth,
height=1.7in} \epsfig{file=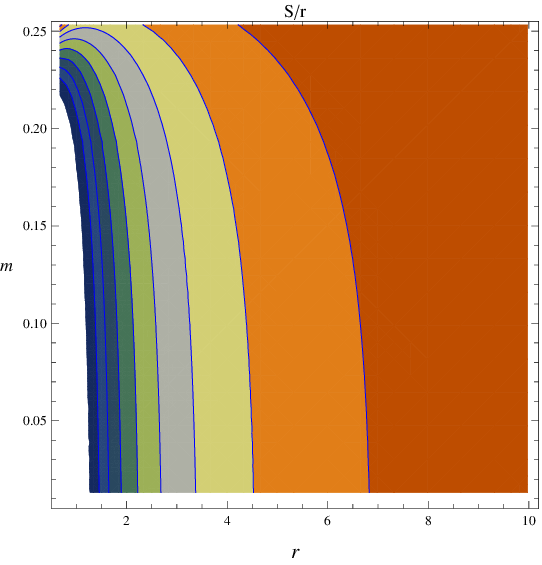, width=.26\linewidth,
height=1.7in}\epsfig{file=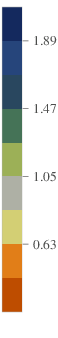, width=.05\linewidth,
height=1.7in} \epsfig{file=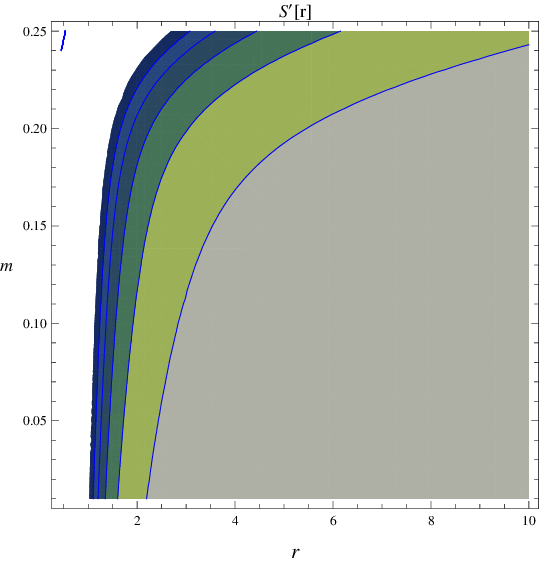, width=.26\linewidth,
height=1.7in}\epsfig{file=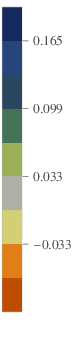, width=.05\linewidth,
height=1.7in}
\caption{\label{F2} shows the shape function properties for model-II with $S(r)$(left), $\frac{S(r)}{r}$ (middle) and $\frac{dS}{dr}$ (right).}
\end{figure}
For the current analysis, the shape function governed by Eq. (\ref{shape2}) should be considered as model-II. Indeed, there is no concept of event horizon in the WH structural geometry. A graphical analysis of both newly calculated shape functions by Eq. (\ref{shape1}) and Eq. (\ref{shape2}) is presented in Fig. (\ref{F1}) and Fig. (\ref{F2}), respectively. It is necessary to mention that both the calculated embedded shape functions are viable and fulfill the required properties for the existence of WH. Both calculated shape functions are observed to be positive, with increasing behavior throughout the configurations. Flaring out and flatness properties are also fulfilled by the particular values of the involved parameters. One can confirm from Fig. (\ref{F1}) and Fig. (\ref{F2}) that the derivative of the shape function concerning radial coordinate remains less than one and the ratio of shape function and radial coordinate approaches zero when radial coordinate approaches to infinity. The satisfying behavior of all the above-mentioned properties shows the viability and physical compatibility of these newly embedded classes of shape functions. The flaring-out condition describes how a WH's mouth expands as it gets away from the throat, which is its narrowest point.    \\

Now, we discuss embedding surface diagrams for two newly calculated WH shape functions with $t = const.$ and $\theta=2\pi$. The embedding surface diagram illustrates how the tunnel joins two distinct locations in spacetime, enabling direct passage between the WH's two mouths. Within the constraint $t = const.$ and $\theta=2\pi$, we have the following relation:
\begin{equation} \label{38}
ds^{2}= r^{2}d\phi^{2}+\left(1-\frac{\mathcal{S}(r)}{r}\right)^{-1}dr^{2}.
\end{equation}
Within the scope of 3-D Euclidean spacetime Eq. (\ref{38}) can be reads as:
\begin{equation} \label{39}
ds^{2}_{\Sigma}= dt^{2}+dr^{2}+r^{2}d\phi^{2}=r^{2}d\phi^{2}+\bigg(1+\bigg(\frac{dZ}{dr}\bigg)^{2}\bigg)dr^{2}.
\end{equation}
On combining Eqs. (\ref{38})-(\ref{39}), we have the following final relation to describe the embedding surface diagram for two newly calculated WH shape functions:
\begin{equation} \label{40}
\frac{dZ}{dr}=\pm \left(\frac{r}{\mathcal{S}(r)}-1\right)^{-1/2}.
\end{equation}
\begin{figure}
\centering \epsfig{file=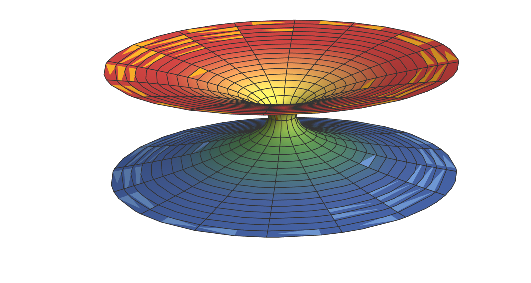, width=.33\linewidth,
height=1.7in} \epsfig{file=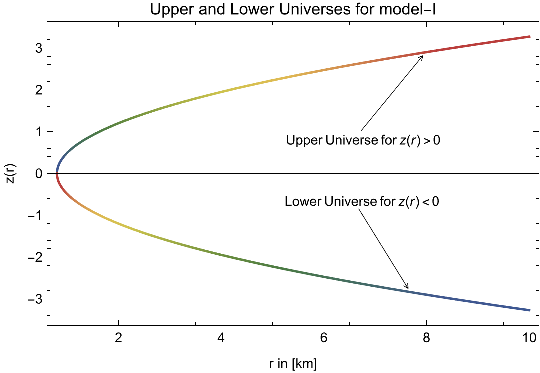, width=.33\linewidth,
height=1.7in} \caption{\label{F3} shows the connection of upper and lower Universes through embedding surface for WH model-I.}
\end{figure}

\begin{figure}
\centering \epsfig{file=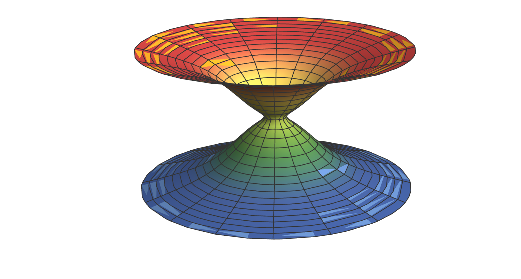, width=.33\linewidth,
height=1.7in} \epsfig{file=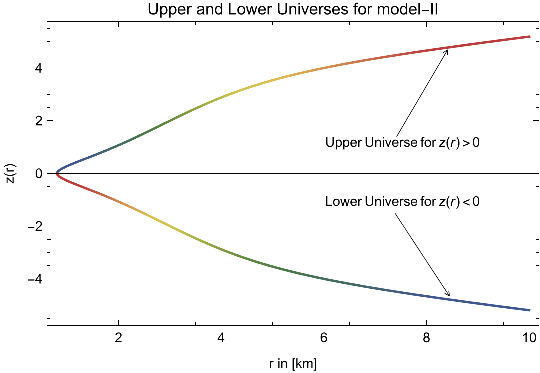, width=.33\linewidth,
height=1.7in} \caption{\label{F4} shows the connection of upper and lower Universes through embedding surface for WH model-II}
\end{figure}
In 1960, Kruskal \cite{jimr3} and Szekeres \cite{jimr4} reported an approach that looked like a maximal extension of the Schwarzschild spacetime in response to the discovery of the Einstein-Rosen bridge \cite{3}. Additionally, they introduced the Kruskal coordinates, a global coordinate system that is essential to understanding the embedding diagram. Fuller and Wheeler \cite{jimr5} demonstrated in 1962 that the Schwarzschild-like WH is not traversable, not even by a photon, by defining its geometry in terms of Kruskal coordinates. In this work, we employ the method first presented by Marolf \cite{jimr7} and Peter and David \cite{jimr6} for creating the embedded diagram of the WH. It is not possible to calculate the exact form of $Z(r)$ from Eq. (\ref{40}) related to two newly calculated shape functions. Due to complications from these shape functions, so we will solve Eq. (\ref{40}) numerically. By fixing the values of involved parameters and WH throat for both WH models, we can create the WH embedded diagrams using numerical techniques, which are presented in Figs. (\ref{F3}) and (\ref{F4}) for model-I and model-II, respectively.  \\

\begin{figure}
\centering \epsfig{file=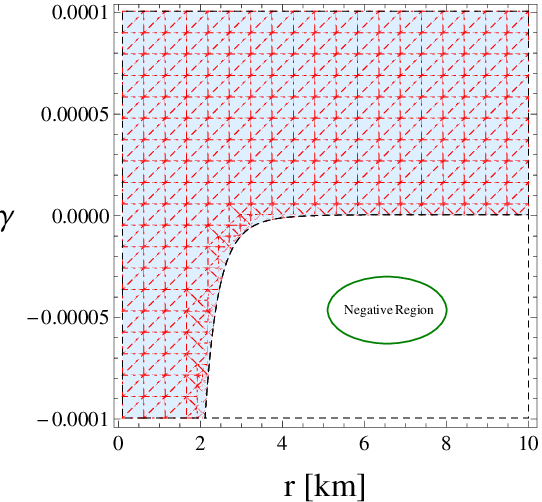, width=.33\linewidth,
height=1.7in} \epsfig{file=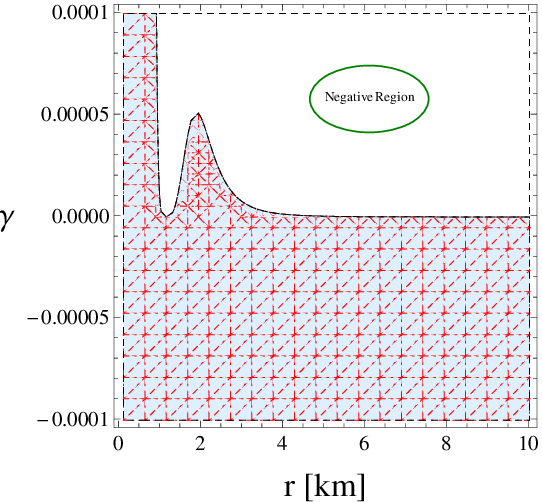, width=.33\linewidth,
height=1.7in}\epsfig{file=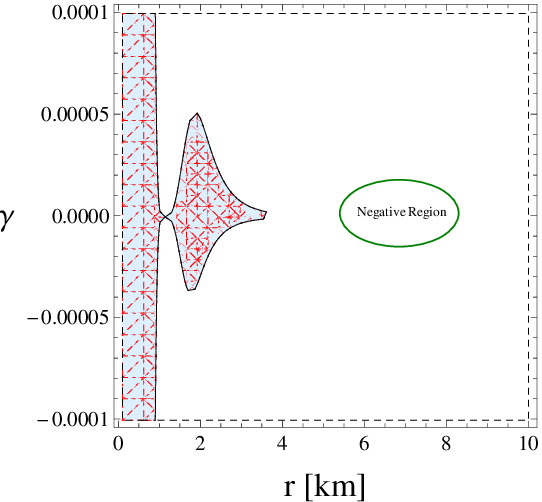, width=.33\linewidth,
height=1.7in}
\centering \epsfig{file=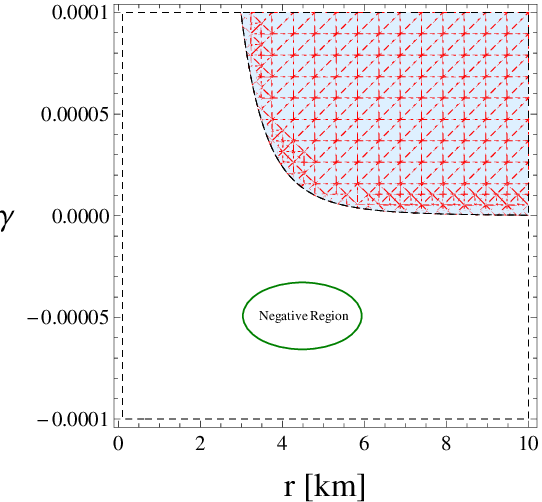, width=.33\linewidth,
height=1.7in} \epsfig{file=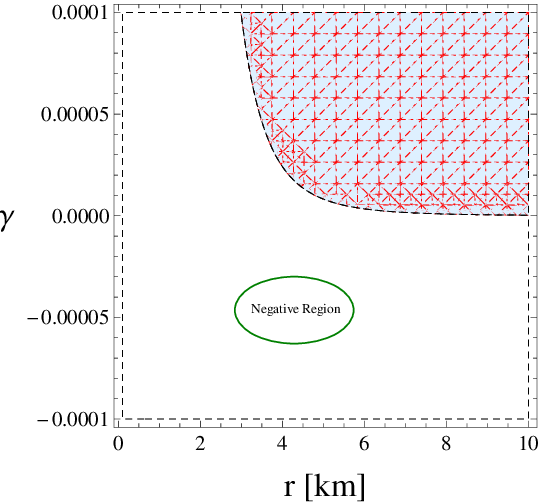, width=.33\linewidth,
height=1.7in}\epsfig{file=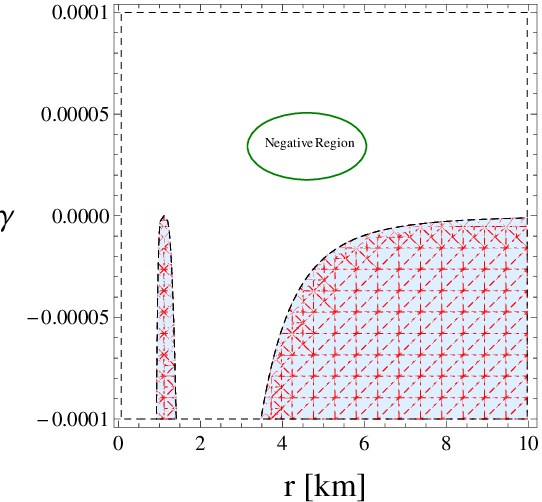, width=.33\linewidth,
height=1.7in}
\caption{\label{F5} shows the valid and negative regions of all the energy conditions for model-I. In the first row $\rho$ (left), $\rho+p_r$ (middle), and $\rho+p_t$ (right) are presnted. In second row, $\rho-p_r$ (left), $\rho-p_t$ (middle), and $\rho+p_r +2p_t$ (right) are given.  }
\end{figure}

\begin{figure}
\centering \epsfig{file=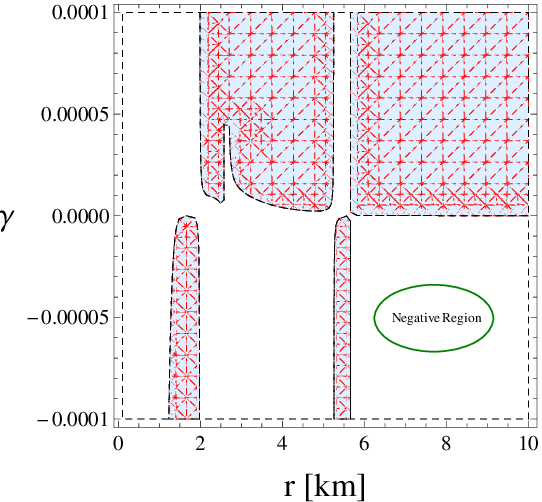, width=.33\linewidth,
height=1.7in} \epsfig{file=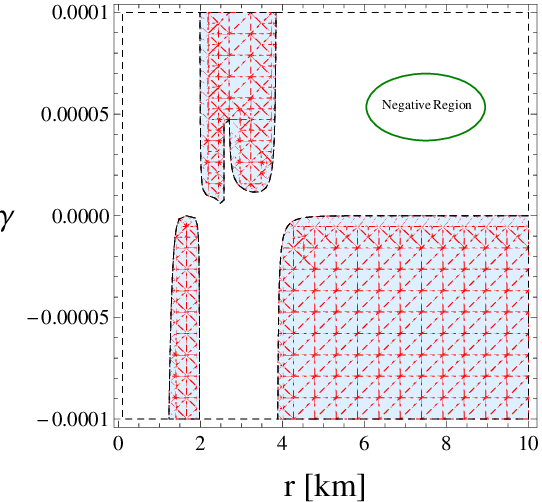, width=.33\linewidth,
height=1.7in}\epsfig{file=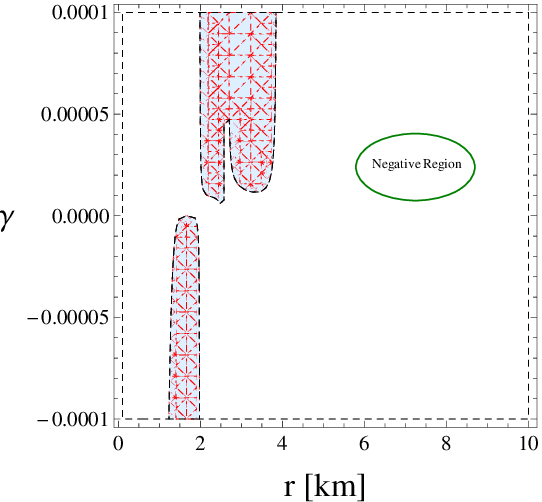, width=.33\linewidth,
height=1.7in}
\centering \epsfig{file=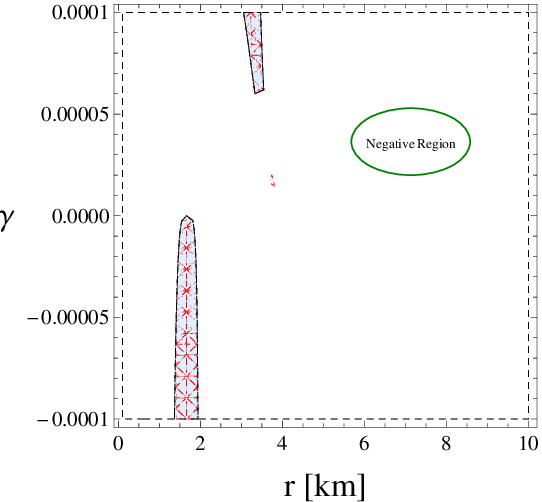, width=.33\linewidth,
height=1.7in} \epsfig{file=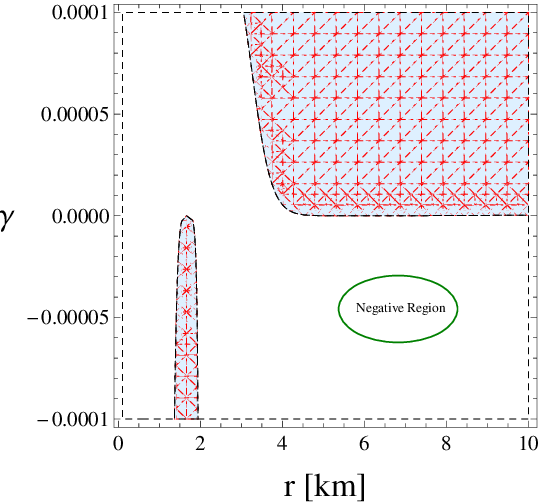, width=.33\linewidth,
height=1.7in}\epsfig{file=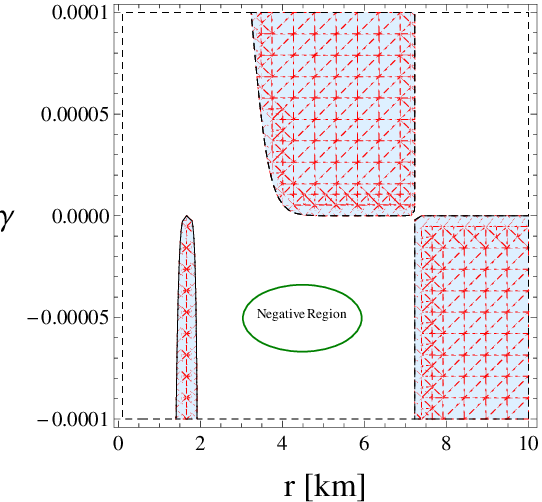, width=.33\linewidth,
height=1.7in}
\caption{\label{F6} shows the valid and negative regions of all the energy conditions for model II. In the first row $\rho$ (left), $\rho+p_r$ (middle), and $\rho+p_t$ (right) are presnted. In second row, $\rho-p_r$ (left), $\rho-p_t$ (middle), and $\rho+p_r +2p_t$ (right) are given. }
\end{figure}

\section{Energy Conditions and their Regional Analysis}

The energy conditions are essential tools for Ricci inverse gravity because they allow us to examine the ordinary and geodesic structure of spacetime thoroughly. The primary energy limitations are the NEC, weak energy condition (WEC), strong energy condition (SEC), and dominant energy condition (DEC). The above conditions are described as
 SEC: $\Leftrightarrow (T_{\epsilon\varepsilon}-\frac{T}{2}g_{\epsilon\varepsilon})X^\epsilon X^\varepsilon\geq 0$,~DEC:$\Leftrightarrow T_{\epsilon\varepsilon}X^\epsilon X^\varepsilon\geq 0$,
 NEC:$\Leftrightarrow T_{\epsilon\varepsilon}\chi^\epsilon \chi^\varepsilon\geq 0$, and ~WEC$\Leftrightarrow T_{\epsilon\varepsilon}X^\epsilon X^\varepsilon\geq 0,$
with $\chi^\epsilon$ is the null vector and $X^\epsilon$ is a time-like vector. For $DEC$, $T_{\epsilon\varepsilon}X^\epsilon$ is not space like. Further, for the principal pressure all the energy conditions are discussed as: SEC: $\Leftrightarrow \forall j (j=r,t),~p_j+\rho\geq 0,~\sum_jp_j+\rho\geq 0,$ DEC: $ \Leftrightarrow\rho\geq 0,~~\forall j,~p_j\epsilon[-\rho,+\rho],$
 NEC: $\Leftrightarrow\forall j,~p_j+\rho\geq 0,$~WEC: $ \Leftrightarrow\rho\geq 0,~~\forall j,~p_j+\rho\geq 0.$ Finally, we have following updated version of all the energy conditions: 
\begin{eqnarray}
\nonumber SEC&:&p_{r}+\rho\geq 0,~~~~~~p_t+\rho\geq 0,~~~~~~~p_r+2p_t+\rho\geq 0,\\
\nonumber DEC&:&\rho\geq 0,~~~~~~~~~~~~-|p_r|+\rho\geq 0,~~~~~~-|p_t|+\rho\geq 0,\\
\nonumber NEC&:&p_r+\rho\geq 0,~~~~~p_t+\rho\geq 0,\\
\nonumber WEC&:&\rho\geq 0,~~~~~~~~~~~~p_r+\rho\geq 0,~~~~~~~p_t+\rho\geq 0.
\end{eqnarray}
The most significant setting for the formation of WH structures is the related evolution of energy density and radial and tangential pressures. The goal of this investigation is to explore the evolution of the energy conditions under the anisotropic source of matter. Fig. (\ref{F5}) and Fig. (\ref{F6}) provide a visual representation of all the energy conditions via regional graphs for both model-I and model-II, respectively. These energy conditions are violated in the context of WHs by the negative energy density needed to keep the WH's throat open. The negative energy required for WH stabilization would require unusual characteristics that are incompatible with general relativity as well as Ricci inverse gravity frameworks. WHs are theorized objects that can exist and provide shortcuts through spacetime to connect two distant points in the universe by violating the energy conditions. Since no actual experiment has seen or verified the existence of such exotic stuff as WHs, it is necessary to mention that the model parameter, i.e., Ricci inverse gravity parameter $\gamma$, has a major influence on all the energy conditions in the current analysis. To explore all the energy conditions, we consider the minimal range of the involved parameter like  $-0.0001\leq \gamma \leq 0.0001$. In this very short range of involved parameters, all the energy conditions are violated in maximum regions; one can confirm the violation of energy condition from Fig. (\ref{F5}) and Fig. (\ref{F6}). The presence of exotic matter is confirmed, which is the requirement for the creation of these two newly calculated WH solutions due to the violation of energy conditions, specifically the violation of NEC. This exotic stuff is thought to contribute to the stability and traversability of WHs by counteracting the gravitational collapse caused by ordinary matter. The study of energy violation in the context of WHs sheds light on these structures. All the energy conditions from Fig. (\ref{F5}) and Fig. (\ref{F6}) for model-I and model-II are supportive of the existence of these two new WH solutions in the background of Ricci inverse gravity.  

\begin{figure}
\centering \epsfig{file=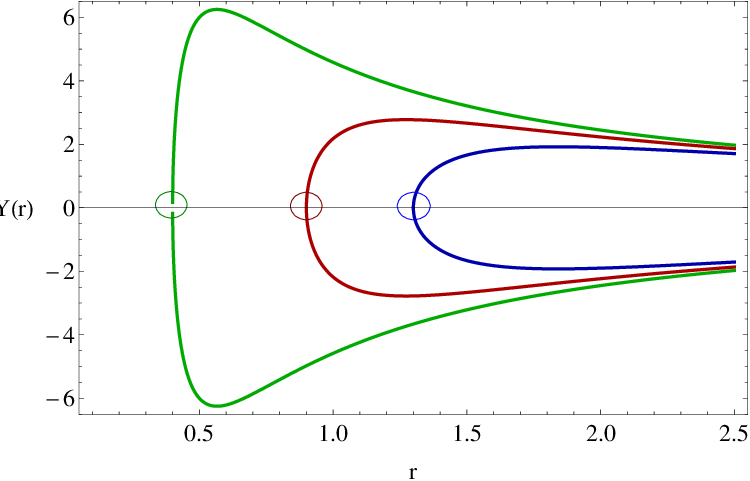, width=.48\linewidth,
height=1.7in} \epsfig{file=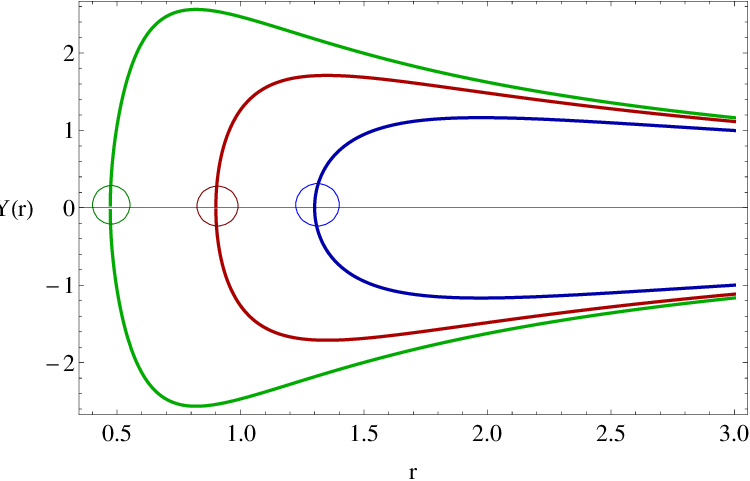, width=.48\linewidth,
height=1.7in}
\caption{\label{F7} shows the allowed-able region for both models with $\left(1-\frac{X(r)}{r}\right)>$ (shaded) and $\left(1-\frac{X(r)}{r}\right)<0$ (un-shaded).}
\end{figure}
\section{Red-Blue Shifts Function}

In this segment, we shall discuss about the red-blue changes of the light coming from the test particles moving around the WH throat in the background of two newly embedded WH solutions. The distance between the detector and the WH throat can be discussed by using the concept of the red-blue shifts function. The following relation can be used to calculate the redshift function for WH geometry \cite{em4}: 
\begin{equation}\label{45}
Y(r)=\pm\sqrt{-\frac{g_{\phi \phi}}{g_{tt}}}\left(\frac{L}{g_{\phi\phi}}-\frac{kA_{\phi}}{g_{\phi\phi}}\right).
\end{equation}
For the present analysis, $A_{\phi}=0$ due to the embedded form of WH geometry. The WH throat can be verified by using the red-blue shifts function. The observer can observe the WH throat location accurately. From Fig. (\ref{F7}), it can be seen that the indications through a small circle from the function $Y$ are similar to the considered WH throat values for the different values of $\zeta _1$ for model-I, and $\chi _1$ for model-II. The red-blue shift phenomena are caused by the Doppler effect, which occurs when an observer moves relative to a wave source, such as light or sound, around the WH's throat. When a WH throat approaches an observer, the waves are compressed, and the frequency looks greater in the blue shift, whereas when an observer moves away, the waves are stretched out, and the frequency appears lower in the redshift.

\section{Discussion}
We have examined a recently suggested fourth-order novel Ricci-inverse gravity that presents the anti-curvature scalar. We have also explored embedded WH geometry in the context of the Ricci-inverse gravitational framework in this letter. Two different embedded classes of shape functions are calculated for WH geometry. 
Gravitational lensing, exotic matter, time dilation and matter displacement are all examples of phenomena that provide empirical evidence for the geometry of WHs or theoretical information that indicates to the occurrence or the existence of a WH.\\

It has been observed that both of the newly calculated shape functions within the scope of embedded space-time meet the Morris and Throne criteria for satisfying the basic structure of WH geometry. In Fig. (\ref{F1}) and Fig. (\ref{F2}), the graphical behavior of required properties is presented for both model-I and model-II, respectively.  The embedded diagrams for both shape functions have been drawn for the upper and lower Universes in Figs. (\ref{F3}) and (\ref{F4}) for both model-I and model-II, respectively. All the energy conditions have been plotted in Fig. (\ref{F5}) and Fig.(\ref{F6}) for both model-I and model-II, respectively. It has been quite interesting that the creation of WH solutions in Ricci inverse gravity includes some exotic matter. One of the most crucial conditions for WH construction is the presence of exotic matter, which causes an NEC violation and makes the WH possible. For both functions, all of the energy conditions with positive as well as negative regions have been provided in Fig. (\ref{F5}) and Fig.(\ref{F6}). In the last section of this letter, we have provided the red-blue shift function. The red-blue shift phenomenon is caused by the Doppler effect, which occurs when an observer moves relative to a wave source, such as light or sound, around WH's throat. When a WH throat approaches an observer, the waves are compressed, and the frequency looks greater in the blue shift, whereas when an observer moves away, the waves are stretched out, and the frequency appears lower in the redshift. The observer can observe the WH throat location accurately. Interestingly, from Fig. (\ref{F7}), it can be seen that the indications through a small circle from the function $Y$ are similar to the considered WH throat values for the different values of $\zeta _1$ for model-I, and $\chi _1$ for model-II. 

It is interesting to mention that the shape functions, which have been calculated in this letter under the framework of embedded class-I, can be considered in other modified theories of gravity to check the compatibility and acceptability of WH geometry. Further, all the calculated results have good agreement with the recently published work \cite{em5}.

\section*{Appendix}
\begin{small}
\begin{eqnarray*}
\rho _1&&=\frac{3 \mathcal{S}-3 r}{p_1}+\frac{r (r-\mathcal{S})}{r \left(\mathcal{S}'-r \Phi'\right)+\mathcal{S} \left(r \Phi'+1\right)},\\
\rho _2&&=\frac{r (r-\mathcal{S})}{\left(-2 r^2 \Phi''+r^2 \left(-\Phi'^2\right)+r \Phi'+4\right)r \left(2 r^2 \Phi''+r^2 \Phi'^2-r \Phi' \mathcal{S}'-4 \mathcal{S}'\right)+\mathcal{S} },\\
\rho _3&&=-2 r^4 (r-\mathcal{S})^2 (2 r-\mathcal{S}) \mathcal{S} \Phi'^6-r^3 (r-\mathcal{S}) \Phi'^5 \left(r^3 \left(5 \mathcal{S}'+4\right)+r^2 \mathcal{S} \left(15-26 \mathcal{S}'\right)\right.\\&&\left.+r \mathcal{S}^2 \left(13 \mathcal{S}'-10\right)-\mathcal{S}^3\right),\\
\rho _4&&=r^2 \mathcal{S}^2 \left(58 r^2 \Phi''-6 r \mathcal{S}''+13 \mathcal{S}'^2-44 \mathcal{S}'-170\right)+2 r \mathcal{S}^3 \left(-22 r^2 \Phi''+r \mathcal{S}''+3 \mathcal{S}'\right.\\&&\left.+67\right)+\mathcal{S}^4 \left(11 r^2 \Phi''-35\right)+r^4 \left(3 r^2 \Phi''+10 \mathcal{S}'^2-12 \mathcal{S}'-24\right)+2 r^3 \mathcal{S} \left(-14 r^2 \Phi''\right.\\&&\left.+2 r \mathcal{S}''-13 \mathcal{S}'^2+28 \mathcal{S}'+46\right),\\
\rho _5&&=-6 r^2 \Phi^{(3)} r^2 (r-\mathcal{S})^2-\Phi''^2 \left(r^2 \left(4 r \mathcal{S}''+15 \mathcal{S}'^2-32 \mathcal{S}'+24\right)-2 r \mathcal{S} \left(2 r \mathcal{S}''-\mathcal{S}'+8\right)\right.\\&&\left.+7 \mathcal{S}^2\right)+3 r^2 (r-\mathcal{S})^2 \Phi''^3+2 r (r-\mathcal{S}) \Phi'' \left(r \Phi^{(4)}(r-\mathcal{S})+\Phi^{(3)} \left(r \left(7 \mathcal{S}'-8\right)+\mathcal{S}\right)\right),\\
\rho _6&&=r^4 \left(\mathcal{S}' \left(-64 r^2 \Phi''+26 r \mathcal{S}''+80\right)+4 \left(r^2 \mathcal{S}^{(3)}+3 r \mathcal{S}''-32\right)-3 \mathcal{S}'^3+72 \mathcal{S}'^2\right)\\&&-r^3 \mathcal{S} \left(\mathcal{S}' \left(-264 r^2 \Phi''+60 r \mathcal{S}''+352\right)+2 \left(8 r^3 \Phi^{(3)}+48 r^2 \Phi''+6 r^2 \mathcal{S}^{(3)}+15 r \mathcal{S}''\right.\right.\\&&\left.\left.-280\right)-3 \mathcal{S}'^3+151 \mathcal{S}'^2\right)+r^2 \mathcal{S}^2 \left(\mathcal{S}' \left(-300 r^2 \Phi''+30 r \mathcal{S}''+503\right)+4 \left(10 r^3 \Phi^{(3)}\right.\right.\\&&\left.\left.+34 r^2 \Phi''+3 r^2 \mathcal{S}^{(3)}+12 r \mathcal{S}''-202\right)+71 \mathcal{S}'^2\right)-r \mathcal{S}^3 \left(32 r^3 \Phi^{(3)}-5 \left(20 r^2 \Phi''-43\right) \mathcal{S}'\right.\\&&\left.+20 r^2 \Phi''+4 r^2 \mathcal{S}^{(3)}+26 r \mathcal{S}''-419\right)+\mathcal{S}^4 \left(8 r^3 \Phi^{(3)}-20 r^2 \Phi''-51\right),\\
\rho _7&& =2 r^2 \Phi''^2 \left(r^3 \left(28-33 \mathcal{S}'\right)+r^2 \mathcal{S} \left(78 \mathcal{S}'-67\right)+\mathcal{S}^2 \left(30 r-39 r \mathcal{S}'\right)+3 \mathcal{S}^3\right)-4 r (r-\mathcal{S})\\&& \times \mathcal{S}\left(2 \Phi^{(3)} \left(r^2 \left(3 r \mathcal{S}''+\mathcal{S}'^2+20\right)-r \mathcal{S} \left(3 r \mathcal{S}''+2 \mathcal{S}'+40\right)+21 \mathcal{S}^2\right)-r \Phi^{(4)}(r-\mathcal{S})\right.\\&&\left.\times \mathcal{S} \left(r \left(\mathcal{S}'-4\right)+3 \mathcal{S}\right)\right)+\Phi'' \left(\left(51-56 r^3 \Phi^{(3)}\right) \mathcal{S}^3+r^3 \left(4 \left(12 r^3 \Phi^{(3)}+r^2 \mathcal{S}^{(3)}-r \mathcal{S}''-48\right)\right.\right.\\&&\left.\left.-3 \mathcal{S}'^3+80 \mathcal{S}'^2+\mathcal{S}' \left(30 r \mathcal{S}''+64\right)\right)-r^2 \mathcal{S} \left(160 r^3 \Phi^{(3)}+8 r^2 \mathcal{S}^{(3)}+22 r \mathcal{S}''+71 \mathcal{S}'^2\right.\right.\\&&\left.\left.+6 \mathcal{S}' \left(5 r \mathcal{S}''+48\right)-512\right)+r \mathcal{S}^2 \left(168 r^3 \Phi^{(3)}+4 r^2 \mathcal{S}^{(3)}+26 r \mathcal{S}''+215 \mathcal{S}'-368\right)\right),
\end{eqnarray*}
\end{small}
\begin{small}
\begin{eqnarray*}
\rho _8&&=-24 r^4 \Phi^{(4)}(r)-72 r^3 \Phi^{(3)}+204 r^4 \Phi''^2-6 r^2 \Phi'' \left(10 r \mathcal{S}''+197\right)+\left(48 r^2 \Phi''+34\right) \mathcal{S}'^2-2 \mathcal{S}'\\&&\times\mathcal{S} \left(30 r^3 \Phi^{(3)}+96 r^2 \Phi''+r^2 \mathcal{S}^{(3)}-10 r \mathcal{S}''-173\right)+20 r^2 \mathcal{S}^{(3)}+6 r^2 \mathcal{S}''^2+259 r \mathcal{S}''+808,\\
\rho _9&&=16 r^4 \Phi^{(4)}(r)+60 r^3 \Phi^{(3)}-144 r^4 \Phi''^2+4 r^2 \Phi'' \left(5 r \mathcal{S}''+216\right)+2 \left(10 r^3 \Phi^{(3)}+16 r^2 \Phi''-69\right) \\&& \times\mathcal{S} \mathcal{S}'-6 r^2 \mathcal{S}^{(3)}-93 r \mathcal{S}''-654,\\
\rho _{10}&&=\mathcal{S}'^2 \left(-96 r^2 \Phi''+r \mathcal{S}''-74\right)+\mathcal{S}' \left(56 r^3 \Phi^{(3)}+260 r^2 \Phi''+4 r^2 \mathcal{S}^{(3)}-42 r \mathcal{S}''-272\right)\\&&-2 \left(-8 r^4 \Phi^{(4)}(r)-24 r^3 \Phi^{(3)}+60 r^4 \Phi''^2-4 r^2 \Phi'' \left(7 r \mathcal{S}''+87\right)+11 r^2 \mathcal{S}^{(3)}+6 r^2 \mathcal{S}''^2\right.\\&&\left.+119 r \mathcal{S}''+224\right)+2 \mathcal{S}'^3,\\
\rho _{11}&&=\mathcal{S}'^2 \left(42 r^2 \Phi''-r \mathcal{S}''+40\right)-2 \mathcal{S}' \left(8 r^3 \Phi^{(3)}+44 r^2 \Phi''+r^2 \mathcal{S}^{(3)}-11 r \mathcal{S}''-32\right)\\&&+2 \left(-2 r^4 \Phi^{(4)}(r)-8 r^3 \Phi^{(3)}+12 r^4 \Phi''^2-8 r^2 \Phi'' \left(r \mathcal{S}''+10\right)+4 r^2 \mathcal{S}^{(3)}+3 r^2 \mathcal{S}''^2\right.\\&&\left.+36 r \mathcal{S}''+48\right)-2 \mathcal{S}'^3,\;\;\;\;\;\;\rho _{12}=p_1^4 (r-\mathcal{S})^2,\\
\rho _{13}&&=2 \rho _4 r^2 \Phi'^4+\rho _6 r \Phi'^3+2 \rho _7 r (r-\mathcal{S}) \Phi'+2 \Phi'^2 \left(2 \mathcal{S}^4 \left(-2 r^4 \Phi^{(4)}(r)-10 r^3 \Phi^{(3)}+18 r^4 \Phi''^2\right.\right.\\&&\left.\left.-112 r^2 \Phi''+99\right)+\rho _{11} r^4+\rho _{10} r^3 \mathcal{S}+\rho _8 r^2 \mathcal{S}^2+\rho _9 r \mathcal{S}^3\right)+\rho _3-8 \rho _5 r^2 (r-\mathcal{S})^2,
\end{eqnarray*}
\end{small}
\begin{small}
\begin{eqnarray*}
p_1&&=2 r (r-\mathcal{S}) \Phi''-\Phi' \left(r \left(\mathcal{S}'-4\right)+3 \mathcal{S}\right)+r (r-\mathcal{S}) \Phi'^2,\\
p_2&&=(r-\mathcal{S})^2 \Phi' \left(r^2 (r-\mathcal{S}) \Phi'^3-r \Phi'^2 \left(r \left(3 \mathcal{S}'-4\right)+\mathcal{S}\right)+2 r \left(2 r \Phi^{(3)} (r-\mathcal{S})-\Phi'' \right.\right.\\&&\left.\left.\times\mathcal{S}\left(3 r \mathcal{S}'+\mathcal{S}-4 r\right)\right)+2 \Phi' \left(r \left(3 r^2 \Phi''-r \mathcal{S}''-2 \mathcal{S}'-4\right)+\mathcal{S} \left(6-3 r^2 \Phi''\right)\right)\right),\\
p_3&&=-3 r^3 \Phi'^5+r^2 \Phi'^4 \left(4 r^2 \Phi''+6 r+15\right)+4 \left(-12 r^5 \Phi''^3+(24-5 r) r^3 \Phi''^2+4 \left(6 r^4 \Phi^{(4)}(r)\right.\right.\\&&\left.\left.+9 r^6 \Phi^{(3)^{2}}+74 r^3 \Phi^{(3)}+85\right)-2 r^2 \left(6 r^4 \Phi^{(4)}(r)+20 r^3 \Phi^{(3)}-1\right) \Phi''\right)+r^2 \Phi'^3 \left(4 (r-6) r^2 \Phi^{(3)}\right.\\&&\left.-2 (7 r+58) r \Phi''-3 r+2\right)+r \Phi'^2 \left(-24 r^4 \Phi^{(4)}(r)-4 (r+32) r^3 \Phi^{(3)}+8 (r+15) r^3 \Phi''^2+2 (5 r+51)\right.\\&&\left.\times \mathcal{S} r^2 \Phi''-17 r-256\right)+2 r \Phi' \left(12 r^4 \Phi^{(4)}(r)+4 (17 r+12) r^2 \Phi^{(3)}-4 (r+8) r^3 \Phi''^2+r \left(4 (r+30)\right.\right.\\&&\left.\left.\times \mathcal{S} r^3 \Phi^{(3)}+27 r+520\right) \Phi''+88\right),\\
p_4&&=-r^3 \Phi'^5 \left(\mathcal{S}'+11\right)+2 r^2 \Phi'^4 \left(8 r^2 \Phi''-r^2 \mathcal{S}''+3 r \mathcal{S}''+(3 r+10) \mathcal{S}'+9 r+20\right)\\&&-8 \left(24 r^5 \Phi''^3-9 r^3 \Phi''^2 \left(r^2 \mathcal{S}''-3 r+4\right)-4 \left(9 r^4 \Phi^{(4)}(r)+18 r^6 \Phi^{(3)^{2}}-2 r \left(9 r^3 \Phi^{(3)}+31\right) \mathcal{S}''\right.\right.\\&&\left.\left.+126 r^3 \Phi^{(3)}-6 r^2 \mathcal{S}^{(3)}+119\right)+\left(-r^3 (17 r+12) \Phi''^2-4 \left(3 r^4 \Phi^{(4)}(r)+22 r^3 \Phi^{(3)}+51\right)+r^2 \right.\right.\\&&\left.\left.\left(24 r^3 \Phi^{(3)}+127\right) \Phi''\right) \mathcal{S}'+r^2 \Phi'' \left(24 r^4 \Phi^{(4)}(r)+56 r^3 \Phi^{(3)}-12 r^2 \mathcal{S}^{(3)}+2 r \mathcal{S}''-131\right)\right)\\&&+r^2 \Phi'^3 \left(16 r^3 \Phi^{(3)}-96 r^2 \Phi^{(3)}-\left(2 r (5 r+18) \Phi''+5 (r-4)\right) \mathcal{S}'-2 (23 r+214) r \Phi''+12 r^2 \mathcal{S}^{(3)}\right.\\&&\left.+2 r^2 \mathcal{S}''+62 r \mathcal{S}''-7 r-12\right)-4 r \Phi'^2 \left(24 r^4 \Phi^{(4)}(r)+3 r^4 \Phi^{(3)}+128 r^3 \Phi^{(3)}-8 (r+15) r^3 \Phi''^2\right.\\&&\left.+r^2 \Phi'' \left((r+24) r \mathcal{S}''-7 r-68\right)+\left(r^4 \Phi^{(3)}-(3 r+34) r^2 \Phi''+16 r+88\right) \mathcal{S}'+3 r^3 \mathcal{S}^{(3)}\right.\\&&\left.-12 r^2 \mathcal{S}^{(3)}+15 r^2 \mathcal{S}''-46 r \mathcal{S}''+r+168\right)+2 r \Phi' \left(-4 (3 r+8) r^3 \Phi''^2-4 \left(-9 r^4 \Phi^{(4)}(r)+6 r \right.\right.\\&&\left.\left.\times\mathcal{S}\left(3 r^3 \Phi^{(3)}+16\right) \mathcal{S}''-6 (11 r+6) r^2 \Phi^{(3)}+12 r^2 \mathcal{S}^{(3)}-95\right)+r \Phi'' \left(16 (r+30) r^3 \Phi^{(3)}+12 r^3\right.\right.\\&&\left.\left.\times\mathcal{S} \mathcal{S}^{(3)}+2 (7 r-144) r \mathcal{S}''+127 r+1776\right)+\left(-4 (r+24) r^3 \Phi''^2-19 (r-16) r \Phi''+4 \left(3 r^4 \Phi^{(4)}(r)\right.\right.\right.\\&&\left.\left.\left.+2 (r+6) r^2 \Phi^{(3)}-7\right)\right) \mathcal{S}'\right),
\end{eqnarray*}
\end{small}
\begin{small}
\begin{eqnarray*}
p_5&&=-3 r^3 \left(\mathcal{S}'+5\right) \Phi'^5+3 r^2 \left(8 \Phi'' r^2-2 \mathcal{S}'' r^2+6 \mathcal{S}'' r+6 r+3 \mathcal{S}'^2+2 (3 r+7) \mathcal{S}'+13\right)\\&&\times\mathcal{S} \Phi'^4+r^2 \left(-(r-42) \mathcal{S}'^2-\left(6 (5 r+18) \Phi'' r-2 (r+21) \mathcal{S}'' r+13 r+24\right) \mathcal{S}'+2 \left(12 \Phi^{(3)} r^3\right.\right.\\&&\left.\left.-72 \Phi^{(3)} r^2+18 \mathcal{S}^{(3)} r^2-3 (9 r+98) \Phi'' r+2 (r+36) \mathcal{S}'' r-2 r-3\right)\right) \Phi'^3-2 r \left(-24 (r+15)\right.\\&&\left. \times\mathcal{S} \Phi''^2 r^3+\Phi'' \left(6 (r+24) \mathcal{S}'' r-13 r-87\right) r^2+\mathcal{S}'^2 \left(37-(r-15) r \Phi''\right) r+2 \mathcal{S}' \left(3 \Phi^{(3)} r^4 \right.\right.\\&&\left.\left. +3 \mathcal{S}^{(3)} r^3-(8 r+117) \Phi'' r^2+3 (3 r-14) \mathcal{S}'' r+11 r+264\right)+2 \left(36 \Phi^{(4)}r^4-9 \mathcal{S}''^2 r^3+3 (r+64) \right.\right.\\&&\left.\left. \times\mathcal{S} \Phi^{(3)} r^3+6 \mathcal{S}^{(3)} r^3-36 \mathcal{S}^{(3)} r^2+12 (3 r-8) \mathcal{S}'' r-4 r+120\right)\right) \Phi'^2+2 r \left(\left(-12 (r+24) \Phi''^2 r^3\right.\right.\\&&\left.\left. +\Phi'' \left(66 \mathcal{S}'' r^2-43 r+1344\right) r+4 \left(9 \Phi^{(4)}r^4+36 (r+1) \Phi^{(3)} r^2-12 \mathcal{S}^{(3)} r^2-72 \mathcal{S}'' r+133\right)\right) \right.\\&&\left. \mathcal{S}'-\mathcal{S}'^2 \left(60 \Phi^{(3)} r^3+(7 r+216) \Phi'' r+308\right)+4 \left(-54 \mathcal{S}'' \Phi^{(3)} r^4+9 \Phi^{(4)}r^4-3 (r-8) \right.\right.\\&&\left.\left. \times\mathcal{S} \Phi''^2 r^3+81 \Phi^{(3)} r^3+36 \mathcal{S}''^2 r^2+36 \Phi^{(3)} r^2-24 \mathcal{S}^{(3)} r^2-216 \mathcal{S}'' r+\Phi'' \left(6 (r+30) \Phi^{(3)} r^3 \right.\right.\right.\\&&\left.\left. \left.+9 \mathcal{S}^{(3)} r^3-6 (r+36) \mathcal{S}'' r+53 r+498\right) r+76\right)\right) \Phi'+4 \left(-72 \Phi''^3 r^5+3 \Phi''^2 \left(18 \mathcal{S}'' r^2-31 r\right.\right.\\&&\left.\left.+24\right) r^3-4 \Phi'' \left(18 \Phi^{(4)}r^4+24 \Phi^{(3)} r^3-18 \mathcal{S}^{(3)} r^2+36 \mathcal{S}'' r-149\right) r^2+\mathcal{S}'^2 \left(39 \Phi''^2 r^4\right.\right.\\&&\left.\left. -178 \Phi'' r^2-24 \left(5 \Phi^{(3)} r^3+8\right)\right)+8 \left(18 r^2 \mathcal{S}''^2-2 r \left(27 \Phi^{(3)} r^3+68\right) \mathcal{S}''+3 \left(9 \Phi^{(3)^{2}} r^6\right.\right.\right.\\&&\left.\left.\left. +3 \Phi^{(4)}r^4+47 \Phi^{(3)} r^3-4 \mathcal{S}^{(3)} r^2+26\right)\right)+2 \mathcal{S}' \left(12 (r+3) \Phi''^2 r^3+\Phi'' \left(-72 \Phi^{(3)} r^3\right.\right.\right.\\&&\left.\left.\left.+66 \mathcal{S}'' r-203\right) r^2+4 \left(9 \Phi^{(4)}r^4+96 \Phi^{(3)} r^3-6 \mathcal{S}^{(3)} r^2-50 \mathcal{S}'' r+201\right)\right)\right),
\end{eqnarray*}
\end{small}
\begin{small}
\begin{eqnarray*}
p_6&&=-r^3 \left(\mathcal{S}'+2\right) \Phi'^5+r^2 \left(4 \Phi'' r^2-2 \mathcal{S}'' r^2+6 \mathcal{S}'' r+9 \mathcal{S}'^2+(6 r+2) \mathcal{S}'+4\right) \Phi'^4\\&&+r^2 \left((r+24) \mathcal{S}'^3-2 (2 r+15) \mathcal{S}'^2+\left(-2 r (5 r+18) \Phi''+2 r (r+21) \mathcal{S}''+8\right) \mathcal{S}'+4 r \right.\\&&\left.\times\mathcal{S} \left(-(r+20) \Phi''+5 \mathcal{S}''+r \left((r-6) \Phi^{(3)}+3 \mathcal{S}^{(3)}\right)\right)\right) \Phi'^3+r \left(3 r \mathcal{S}'^4+(96-36 r) \mathcal{S}'^3\right.\\&&\left. +2 \left((r-15) \Phi'' r^2+4 \left(3 \mathcal{S}'' r^2+2 r-36\right)\right) \mathcal{S}'^2-4 \left(\Phi^{(3)} r^4+3 \mathcal{S}^{(3)} r^3-(2 r+49) \Phi'' r^2\right.\right.\\&&\left.\left.+21 (r-2) \mathcal{S}'' r+16\right) \mathcal{S}'+4 r \left(2 r^2 (r+15) \Phi''^2-r \left(r (r+24) \mathcal{S}''+16\right) \Phi''+9 r^2 \mathcal{S}''^2+4 \mathcal{S}''\right.\right.\\&&\left.\left.-2 r \left(3 \Phi^{(4)}r^2+16 \Phi^{(3)} r-6 \mathcal{S}^{(3)}\right)\right)\right) \Phi'^2+2 r \left(12 \mathcal{S}'^4+3 \left(13 r^2 \Phi''-76\right) \mathcal{S}'^3-4 \left(15 \right.\right.\\&&\left.\left.\times\mathcal{S} \Phi^{(3)} r^3+(31 r+54) \Phi'' r-24 \mathcal{S}'' r-76\right) \mathcal{S}'^2+2 r \left(-2 r^2 (r+24) \Phi''^2+\left(33 \mathcal{S}'' r^2+56 r+368\right)\right.\right.\\&&\left. \left. \times\mathcal{S} \Phi''-240 \mathcal{S}''+2 r \left(4 (8 r+3) \Phi^{(3)}+3 \left(r^2 \Phi^{(4)}(r)-4 \mathcal{S}^{(3)}\right)\right)\right) \mathcal{S}'+4 r^2 \left(16 r \Phi''^2+\left(r^2 \right.\right.\right.\\&&\left.\left.\left. \times\mathcal{S} \left((r+30) \Phi^{(3)}+3 \mathcal{S}^{(3)}\right)-(13 r+72) \mathcal{S}''\right) \Phi''+18 \mathcal{S}'' \left(2 \mathcal{S}''-r^2 \Phi^{(3)}\right)\right)\right) \Phi'+4 \\&&\times\mathcal{S}\left(12 \mathcal{S}'^4+\left(78 r^2 \Phi''-296\right) \mathcal{S}'^3+\left(39 \Phi''^2 r^4-412 \Phi'' r^2+24 \left(-5 \Phi^{(3)} r^3+4 \mathcal{S}'' r+26\right)\right) \right.\\&& \left.\times\mathcal{S} \mathcal{S}'^2-4 r \left(r^2 (11 r-6) \Phi''^2+3 r \left(4 \Phi^{(3)} r^3-11 \mathcal{S}'' r-28\right) \Phi''+2 \left(-3 \Phi^{(4)}r^3-52 \Phi^{(3)} r^2 \right.\right.\right.\\&&\left.\left.\left. +6 \mathcal{S}^{(3)} r+74 \mathcal{S}''\right)\right) \mathcal{S}'-2 r^2 \left(6 \Phi''^3 r^3-9 \Phi''^2 \mathcal{S}'' r^3+2 \Phi'' \left(3 \Phi^{(4)}r^3-2 \Phi^{(3)} r^2\right.\right.\right.\\&&\left.\left.\left.-6 \mathcal{S}^{(3)} r+34 \mathcal{S}''\right) r-18 \left(r^2 \Phi^{(3)}-2 \mathcal{S}''\right)^2\right)\right),\\
p_7&&=-2 (r-40) r^2 \Phi''^2+2 \left(-6 \left(9 r^3 \Phi^{(3)}+20\right) \mathcal{S}''+r \left(3 \left(r^2 \Phi^{(4)}(r)-4 \mathcal{S}^{(3)}\right)+4 (8 r+3) \Phi^{(3)}\right) \right.\\&&\left.+72 r \mathcal{S}''^2\right)+\Phi'' \left(8 (r+30) r^3 \Phi^{(3)}+18 r^3 \mathcal{S}^{(3)}-9 (5 r+48) r \mathcal{S}''+56 r+368\right),\\
p_8&&=3 \left(13 r^2 \Phi''-60\right) \mathcal{S}'^3-\mathcal{S}'^2 \left(120 r^3 \Phi^{(3)}+(131 r+432) r \Phi''-96 r \mathcal{S}''+76\right)-4 \mathcal{S}' \\&& \times\mathcal{S} \left(-9 r^4 \Phi^{(4)}(r)-66 r^3 \Phi^{(3)}-36 r^2 \Phi^{(3)}+3 (r+24) r^3 \Phi''^2-r \Phi'' \left(33 r^2 \mathcal{S}''+22 r+444\right)\right.\\&&\left.+24 r^2 \mathcal{S}^{(3)}+192 r \mathcal{S}''-152\right)+2 p_7 r,
\end{eqnarray*}
\end{small}
\begin{small}
\begin{eqnarray*}
p_9&&=2 p_8 r \Phi'-3 r^3 \Phi'^5 \left(\mathcal{S}'+3\right)+2 r^2 \Phi'^4 \left(8 r^2 \Phi''-3 r^2 \mathcal{S}''+9 r \mathcal{S}''+9 \mathcal{S}'^2+3 (3 r+4) \mathcal{S}'\right.\\&&\left.+3 r+9\right)+8 \left(\left(39 r^2 \Phi''-124\right) \mathcal{S}'^3+\mathcal{S}'^2 \left(-120 r^3 \Phi^{(3)}+39 r^4 \Phi''^2-295 r^2 \Phi''+48 r \mathcal{S}''+180\right)\right.\\&&\left.+r \left(-24 r^4 \Phi''^3+r^2 \Phi''^2 \left(27 r^2 \mathcal{S}''-22 r+12\right)+4 \left(-2 \left(27 r^3 \Phi^{(3)}+37\right) \mathcal{S}''+r \left(3 r^2 \Phi^{(4)}(r)\right.\right.\right.\right.\\&&\left.\left.\left.\left.+18 r^4 \Phi^{(3)^{2}}+52 r \Phi^{(3)}-6 \mathcal{S}^{(3)}\right)+36 r \mathcal{S}''^2\right)-2 r \Phi'' \left(12 r^4 \Phi^{(4)}(r)+4 r^3 \Phi^{(3)}-18 r^2 \mathcal{S}^{(3)}\right.\right.\right.\\&&\left.\left.\left.+69 r \mathcal{S}''-84\right)\right)+\mathcal{S}' \left(36 r^4 \Phi^{(4)}(r)+504 r^3 \Phi^{(3)}-9 (3 r-4) r^3 \Phi''^2+4 r^2 \Phi'' \left(-18 r^3 \Phi^{(3)}+33 r \right.\right.\right.\\&&\left.\left.\left. \times\mathcal{S} \mathcal{S}''+23\right)-48 r^2 \mathcal{S}^{(3)}-496 r \mathcal{S}''+624\right)\right)+r^2 \Phi'^3 \left(-2 \mathcal{S}' \left(3 (5 r+18) r \Phi''-2 (r+21) r \mathcal{S}'' \right.\right.\\&&\left.\left. +4 r+18\right)+2 \left(8 r^3 \Phi^{(3)}-48 r^2 \Phi^{(3)}-(13 r+178) r \Phi''+18 r^2 \mathcal{S}^{(3)}+(r+51) r \mathcal{S}''+4\right)+(r+24)\right.\\&&\left. \times\mathcal{S} \mathcal{S}'^3+(12-5 r) \mathcal{S}'^2\right)-4 r \Phi'^2 \left(24 r^4 \Phi^{(4)}(r)+r^4 \Phi^{(3)}+128 r^3 \Phi^{(3)}-8 (r+15) r^3 \Phi''^2+r^2 \Phi'' \right.\\&&\left. \times\mathcal{S} \left(3 (r+24) r \mathcal{S}''-2 r+15\right)+\mathcal{S}'^2 \left(-(r-15) r^2 \Phi''-6 r^2 \mathcal{S}''+19 r+72\right)+\mathcal{S}' \left(3 r^4 \Phi^{(3)}\right.\right.\\&&\left.\left.-(7 r+132) r^2 \Phi''+6 r^3 \mathcal{S}^{(3)}+6 (5 r-14) r \mathcal{S}''-8 r+192\right)+3 r^3 \mathcal{S}^{(3)}-18 r^3 \mathcal{S}''^2-36 r^2 \mathcal{S}^{(3)}\right.\\&&\left.+21 r^2 \mathcal{S}''-54 r \mathcal{S}''+6 (r-4) \mathcal{S}'^3+16\right),\\
p_{10}&&=r \left(-2 r^2 \Phi''+r \Phi' \mathcal{S}'-r \Phi'^2+4 \mathcal{S}'\right)+\mathcal{S} \left(2 r^2 \Phi''+r \Phi'^2-r \Phi'-4\right),\\
p_{11}&&=\frac{(r-\mathcal{S})^2 \left(2 r \Phi''-\frac{8 \Phi r^5}{(r-\mathcal{S}) \left(r \left(\mathcal{S}'-r \Phi'\right)+\mathcal{S} \left(r \Phi'+1\right)\right)}+\frac{\left(r \Phi'+2\right) \left(r^2 \Phi'-r \mathcal{S} \Phi'-r \mathcal{S}'+\mathcal{S}\right)}{r (r-\mathcal{S})}\right)}{r^3},\\
p_{12}&&=\frac{r-\mathcal{S}}{r^2 \left(-\Phi'\right)+r \mathcal{S} \Phi'+r \mathcal{S}'+\mathcal{S}},\\
p_{13}&&=\frac{\frac{\mathcal{S}-r}{p_1}-\frac{r (r-\mathcal{S})}{r \left(2 r^2 \Phi''+r^2 \Phi'^2-r \Phi' \mathcal{S}'-4 \mathcal{S}'\right)+\mathcal{S} \left(-2 r^2 \Phi''+r^2 \left(-\Phi'^2\right)+r \Phi'+4\right)}}{r},\\
p_{14}&&=r^2 (r-\mathcal{S}) \Phi'^3-r \Phi'^2 \left(r \left(3 \mathcal{S}'-4\right)+\mathcal{S}\right)+2 r \left(2 r \Phi^{(3)} (r-\mathcal{S})-\Phi'' \left(3 r \mathcal{S}'+\mathcal{S}\right.\right.\\&&\left.\left.-4 r\right)\right)+2 \Phi' \left(r \left(3 r^2 \Phi''-r \mathcal{S}''-2 \mathcal{S}'-4\right)+\mathcal{S} \left(6-3 r^2 \Phi''\right)\right),\\
p_{15}&&=-6 r^3 \Phi^{(3)}+36 r^4 \Phi''^2-36 r^3 \Phi'' \mathcal{S}''+\left(4-4 r^2 \Phi''\right) \mathcal{S}'^2-71 r^2 \Phi''-r^3 \Phi'^3 \left(\mathcal{S}'+2\right)\\&&+\mathcal{S}' \left(6 r^3 \Phi^{(3)}+8 r^2 \Phi''-2 r^2 \mathcal{S}^{(3)}-2 r \mathcal{S}''+8\right)+r^2 \Phi'^2 \left(3 r^2 \Phi''+5 r \mathcal{S}''+6 \mathcal{S}'^2\right.\\&&\left.+18 \mathcal{S}'-8\right)+r \Phi' \left(\mathcal{S}' \left(-24 r^2 \Phi''+10 r \mathcal{S}''+35\right)-2 \left(6 r^3 \Phi^{(3)}+16 r^2 \Phi''-3 r^2 \mathcal{S}^{(3)}\right.\right.\\&&\left.\left.-14 r \mathcal{S}''+6\right)\right)+4 r^2 \mathcal{S}^{(3)}+6 r^2 \mathcal{S}''^2+47 r \mathcal{S}''+12,\\
p_{16}&&=r^3 \mathcal{S} \left(\left(10-9 r^2 \Phi''\right) \mathcal{S}'^2-r^3 \Phi'^3 \left(2 \mathcal{S}'+1\right)+2 r \left(r \left(\mathcal{S}^{(3)}-r \Phi^{(3)}\right)+12 r^3 \Phi''^2-6 r \Phi''\right.\right.\\&&\left.\left.\times\mathcal{S} \left(3 r \mathcal{S}''+2\right)+6 r \mathcal{S}''^2+11 \mathcal{S}''\right)+\mathcal{S}' \left(6 r^3 \Phi^{(3)}+4 r^2 \Phi''-4 r^2 \mathcal{S}^{(3)}+8\right)+r^2 \Phi'^2 \left(3 r^2 \Phi''\right.\right.\\&&\left.\left.+4 r \mathcal{S}''+12 \mathcal{S}'^2+2 \mathcal{S}'-6\right)+r \Phi' \left(-4 \mathcal{S}' \left(6 r^2 \Phi''-5 r \mathcal{S}''-3\right)-2 \left(4 r^3 \Phi^{(3)}+5 r^2 \Phi''\right.\right.\right.\\&&\left.\left.\left.-3 r^2 \mathcal{S}^{(3)}-5 r \mathcal{S}''+6\right)+5 \mathcal{S}'^2\right)\right)+r^4 \left(-\left(2 r \mathcal{S}' \left(r^2 \Phi^{(3)}-r \mathcal{S}^{(3)}+\mathcal{S}''\right)+6 r^2 \left(\mathcal{S}''\right.\right.\right.\\&&\left.\left.\left.-r \Phi''\right)^2+\mathcal{S}'^2 \left(-5 r^2 \Phi''+r \mathcal{S}''+4\right)-r^3 \Phi'^3 \mathcal{S}'+r^2 \Phi'^2 \left(r \left(r \Phi''+\mathcal{S}''\right)+6 \mathcal{S}'^2\right.\right.\right.\\&&\left.\left.\left.-4 \mathcal{S}'\right)+r \Phi' \left(-2 r \left(r^2 \Phi^{(3)}-r \mathcal{S}^{(3)}+\mathcal{S}''\right)-2 \mathcal{S}' \left(4 r^2 \Phi''-5 r \mathcal{S}''+2\right)+\mathcal{S}'^3\right.\right.\right.\\&&\left.\left.\left.+4 \mathcal{S}'^2\right)\right)\right)+r \mathcal{S}^3 \left(2 \left(r^3 \Phi^{(3)} \left(\mathcal{S}'-3\right)+r^2 \mathcal{S}^{(3)}+13 r \mathcal{S}''+15\right)+24 r^4 \Phi''^2+r^2 \Phi'' \right.\\&&\left.\times\mathcal{S}\left(-12 r \mathcal{S}''+4 \mathcal{S}'-71\right)-r^3 \Phi'^3+r^2 \Phi'^2 \left(r^2 \Phi''+2 r \mathcal{S}''+12 \mathcal{S}'+4\right)+r \Phi' \left(-8 r^3 \Phi^{(3)}\right.\right.\\&&\left.\left.+\left(20-8 r^2 \Phi''\right) \mathcal{S}'-34 r^2 \Phi''+2 r^2 \mathcal{S}^{(3)}+16 r \mathcal{S}''+19\right)\right)+2 \mathcal{S}^4 \left(r^3 \Phi^{(3)}-3 r^4 \Phi''^2\right.\\&&\left.+12 r^2 \Phi''-3 r^2 \Phi'^2+r \Phi' \left(r^3 \Phi^{(3)}+6 r^2 \Phi''-10\right)-10\right)-p_{15} r^2 \mathcal{S}^2,\\
p_{17}&&=\frac{p_{16}}{\left(r \left(\mathcal{S}'-r \Phi'\right)+\mathcal{S} \left(r \Phi'+1\right)\right)^4}+\frac{2 p_{14} (r-\mathcal{S})^2 \Phi'}{p_1^3 r}.
\end{eqnarray*}
\end{small}

\end{document}